\documentclass[]{interact}

\usepackage{epstopdf}
\usepackage[caption=false]{subfig}
\usepackage{amsmath}
\usepackage[numbers,sort&compress]{natbib}

\bibpunct[, ]{[}{]}{,}{n}{,}{,}

\theoremstyle{plain}

\theoremstyle{definition}

\theoremstyle{remark}

\begin{document}


\title{Dimensional scaling of flame propagation in discrete~particulate~clouds}

\author{
\name{Fredric Y.K. Lam\textsuperscript{a}, XiaoCheng Mi\textsuperscript{b}, Andrew J. Higgins\textsuperscript{c}}
\thanks{Corresponding author: A.J. Higgins. Email: andrew.higgins@mcgill.ca}
\affil{\textsuperscript{a}Institute for Computational and Mathematical Engineering, Stanford University, Stanford, CA 94305, USA\\
\textsuperscript{b}Laboratory for Scientific Computing, Cavendish Laboratory, Department of Physics, University of Cambridge, Cambridge CB3 0HE, UK\\ \textsuperscript{c}Department of Mechanical Engineering, McGill University, Montreal, QC H3A 0C3, Canada}
}

\maketitle

\begin{abstract}
The critical dimension necessary for a flame to propagate in suspensions of fuel particles in oxidizer is studied analytically and numerically.  Two types of models are considered:  First, a continuum model, wherein the individual particulate sources are not resolved and the heat release is assumed spatially uniform, is solved via conventional finite difference techniques.  Second, a discrete source model, wherein the heat diffusion from individual sources is modeled via superposition of the Green's function of each source, is employed to examine the influence of the random, discrete nature of the media.  Heat transfer to cold, isothermal walls and to a layer of inert gas surrounding the reactive medium are considered as the loss mechanisms.  Both cylindrical and rectangular (slab) geometries of the reactive medium are considered, and the flame speed is measured as a function of the diameter and thickness of the domains, respectively. In the continuum model with inert gas confinement, a universal scaling of critical diameter to critical thickness near 2:1 is found. In the discrete source model, as the time scale of heat release of the sources is made small compared to the interparticle diffusion time, the geometric scaling between cylinders and slabs exhibits values greater than 2:1. The ability of the flame in the discrete regime to propagate in thinner slabs than predicted by continuum scaling is attributed to the flame being able to exploit local fluctuations in concentration across the slab to sustain propagation. As the heat release time of the sources is increased, the discrete source model reverts back to results consistent with the continuum model. Implications of these results for experiments are discussed.
\end{abstract}

\begin{keywords}
Metal particle combustion; flame propagation limit; percolation; heterogeneous combustion; critical phenomena; statistical phenomena; geometric scaling
\end{keywords}

\section{Introduction}

The examination of how flames respond to heat losses reveals considerable insight into flame structure and dynamics.  It could be argued that modern combustion science began in 1815 with Sir Humphry Davy's development of the safety lamp. In addition to its practical application in preventing mine explosions, the demonstration that a wire mesh can quench a flame made clear that it is the competition between heat generation and heat loss that determines criticality in flame propagation.~\cite{Davy1816} The seminal analytic treatment of this problem by Spalding showed that quenching corresponds to a criticality in the solution of the flame structure.~\cite{Spalding1957} Modern asymptotic analysis revealed that the decrement in flame speed with increasing heat loss can be described mathematically by an eigenvalue solution for the flame speed, with the critical condition for flame quenching associated with a turning point in the solution.~\cite{Buckmaster1976,Joulin1976} 

Recently, a number of unique features of flame propagation have been experimentally identified in suspensions of reactive metal particles in a gas-phase oxidizing environment, wherein the inherent fluctuations in particle spacing may occur on a scale that is comparable to the flame thickness itself.~\cite{Goroshin2011PRE,Wright2016} The resulting flame propagation is thus dominated by the local heat diffusion among randomly distributed particles rather than a laminar-like reaction front, as in classical flame theory.~\cite{GoroshinLeeShoshin1998,Tang2009CTM,Goroshin2011PRE,Tang2011PRE,Wright2016,Lam2017PRE} The flame propagation behavior identified in such systems has now been recognized as a unique branch of combustion, \emph{discrete combustion}.~\cite{Mukasyan2008} In these earlier studies, the discrete regime of flame propagation has been examined in adiabatic systems (or quasi-adiabatic systems with insignificant heat loss). Further in-depth exploration into the physics of discrete combustion, however, necessitates examining how such flames respond to heat losses.


In this paper, flame quenching is examined in a model system consisting of three dimensional clouds of randomly positioned heat sources and is compared between cylindrical and rectangular slabs of the reactive media. Flame propagation in this system is amenable to description via an analytic construction of superposition of Green's functions representing the sources; however, in practice, a computer is required in order to keep track of the sequence of ignitions of the large number of sources (approaching $10^5$ sources for some of the simulations in this paper). Taking advantage of this modeling approach, Lam~\textit{et al.} previously performed a large number of simulations in randomly generated, two-dimensional clouds of sources with periodic boundary conditions (and thus no losses), and the resulting flame propagations were ensemble-averaged to measure the thickness of the flame front due to roughening as it propagated through the random media.  In that study, it was found that, in the highly discrete regime with a low ignition temperature, the flame front roughening behavior could be described by a power law relation that was shown to belong to  the Kardar-Parisi-Zhang (KPZ) universality class.~\cite{Lam2017PRE} The present study extends these results to three dimensions and includes losses at the periphery of the reactive media, such that critical behavior may be observed. The fact that the resulting phenomenon for the problem under consideration is statistical (meaning, for a given concentration, a flame in a given geometry may propagate or quench depending upon the particular realization of the system) necessitates that ensemble averaging also be employed in the present study, assisted by the use of high performance computing clusters. 
The flame front morphologies, propagation speed, and scaling between critical cloud diameter and thickness resulting from the discrete source model will be compared to the classical solution of a thermal flame described by continuum equations and discussed in this paper.

This paper is organized as follows. In Sect.~2, the models describing a continuous reactive medium (i.e., continuum model) and a cloud of reactive particulates (i.e., discrete source model) are introduced. The heat losses into cold surroundings (either inert gas or isothermal wall) considered in these models are also described in Sect.~2. The results of flame front morphology, flame propagation speed, critical cloud dimension for flame propagation, and the dimensional scaling between the critical cloud dimensions of cylindrical and slab geometries are presented in Sect.~3. The results are discussed and interpreted in Sect.~4. The key findings of this study are summarized in the Conclusion (Sect.~5). Some mathematical details of how the governing equations of the models were derived from the reactive Navier-Stokes equations and how the boundary conditions were implemented in the calculations can be found in the Appendices. 

\section{Model description}
\label{Sec2}

In this study, since the flame propagation behavior resulting from a system that consists of discrete sources will be compared with the classical solution of thermal flames, two different models are considered to treat a continuous reactive medium and a suspension of randomly distributed, point-like sources. These models are referred to as the \emph{continuum model} and \emph{discrete source model}, and introduced in Sect.~\ref{Sec2_1} and Sect.~\ref{Sec2_2}, respectively. The heat losses through the boundaries of the reactive medium are implemented into these models via applying boundary conditions (i.e., isothermal wall or open boundary conditions) to the governing equations of the models. The details of the implementation of boundary conditions can be found in Sect.~\ref{Sec2_3} and Appendix~\ref{SecA2}.

\subsection{Continuum model}
\label{Sec2_1}

In this model, the propagation of thermal flames in a reactive continuum is governed by a two-dimensional heat diffusion equation (in rectangular and axisymmetric coordinates) with a reaction source term and an equation tracking the reaction progress variable. This model is based on the assumption that there is no mass diffusion, i.e., an infinite Lewis number ($\mathrm{Le}=\infty$). The dimensionless form of the governing equations is as follows,
\begin{equation}
\begin{split}
& \frac{\partial \theta}{\partial t} = \frac{\partial^2 \theta}{\partial x^2} + \frac{\partial^2 \theta}{\partial y^2} + R(\theta,C) \;\;\;\;  \mathrm{rectangular}\\
\vspace{0.4mm}
& \frac{\partial \theta}{\partial t} = \frac{\partial^2 \theta}{\partial x^2} + \frac{\partial^2 \theta}{\partial r^2} + \frac{1}{r} \frac{\partial \theta}{\partial r}+ R(\theta,C) \;\;\;\; \mathrm{axisymmetric} \\
\vspace{0.4mm}
& \frac{\partial C}{\partial t} = -R(\theta,C)
\end{split}
\label{Eq1}
\end{equation}
Temperature is non-dimensionalized as $\theta = ( \widetilde{T} - \widetilde{T}_0 ) / ( \widetilde{T}_\mathrm{ad} - \widetilde{T}_0 )$, where $\widetilde{T}$ is the dimensional temperature\footnote{Note that all of the dimensional quantities in this paper are denoted with a tilde ``$\sim$''}, $\widetilde{T}_0$ is the initial temperature, and $\widetilde{T}_\mathrm{ad}$ is the adiabatic flame temperature. $C$ is the dimensionless reaction progress variable, or normalized concentration of reactant, which evolves from $1$ (unreacted) to $0$ (completely reacted). The reaction rate $R$ depends on both $\theta$ and $C$ and is governed by a switch-type reaction model as follows,
\begin{equation}
R(\theta,C) = 
\begin{cases}
1 \;\;\; \theta \geq \theta_\mathrm{ign} \;\;\;\; \mathrm{and} \;\;\;\; C > 0\\
\vspace{0.2mm}
0 \;\;\; \mathrm{otherwise} 
\end{cases}
\label{Eq2}
\end{equation}
When local temperature $\theta$ reaches a prescribed ignition temperature $\theta_\mathrm{ign}$, the exothermic reaction begins. This reaction model is an appropriate approximation for particulate fuels that undergo a transition to diffusion-limited combustion. \cite{SOO2018384}

In the continuum model, time is non-dimensionalized with respect to the characteristic reaction time $\widetilde{t}_\mathrm{r}$, i.e., $t = \widetilde{t}/\widetilde{t}_\mathrm{r}$. The spatial coordinate is $x=\widetilde{x} / \sqrt{\widetilde{\alpha} \widetilde{t}_\mathrm{r}}$ where $\widetilde{\alpha}$ is the thermal diffusivity. Further details of how these dimensionless equations are derived from the dimensional governing equations can be found in Appendix~\ref{SecA1_1} and Ref.~\cite{Mi2016PROCI}.

\begin{figure}[h]
\centerline{\includegraphics[width=\textwidth]{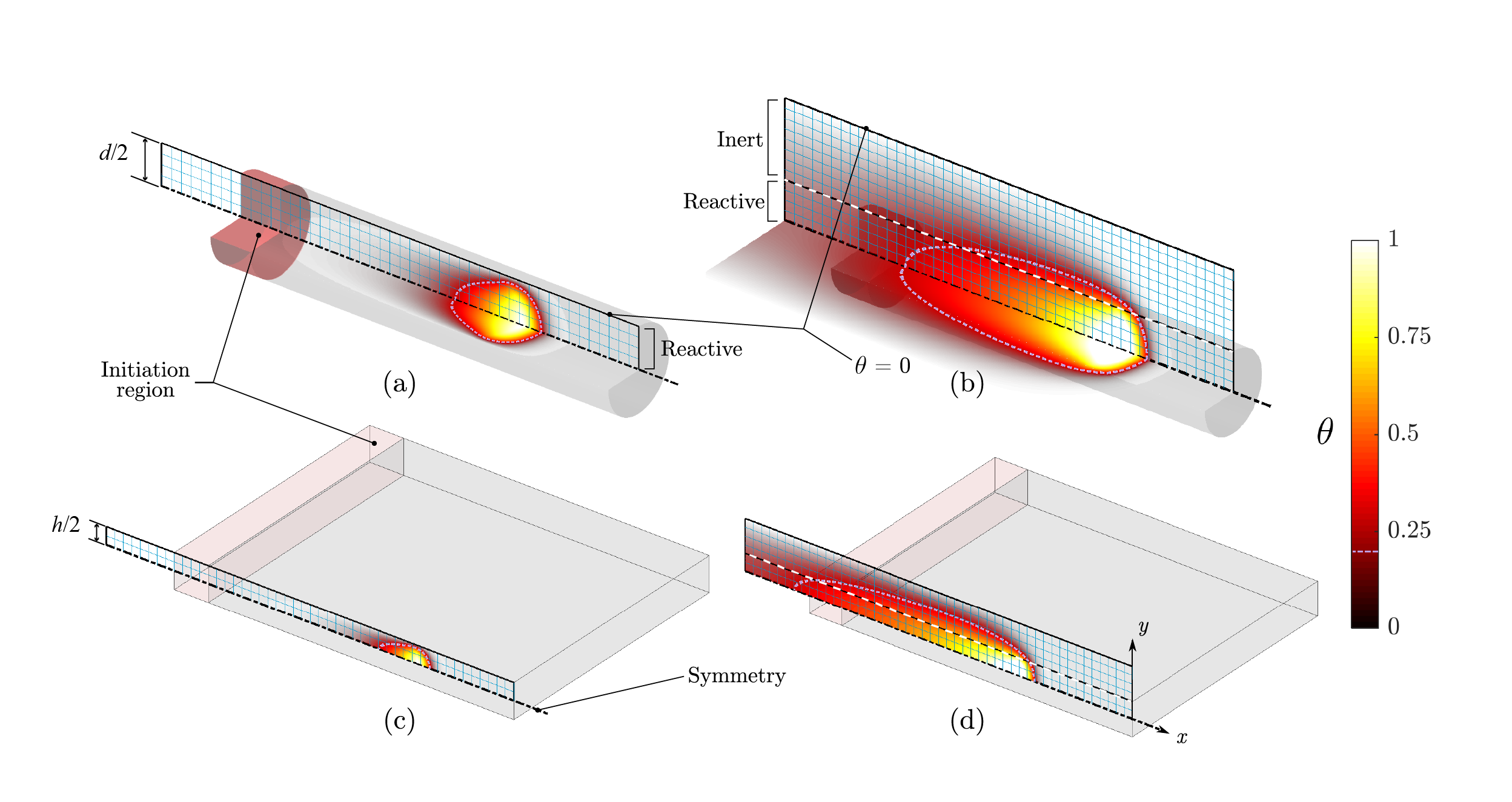}}
\caption{Schematic of continuum model domain. The representative two-dimensional numerical grid is shown with exaggerated grid spacing; solid black lines represent cold wall (Dirichlet) boundary conditions, while dashed-dotted lines represent the axes of symmetry. Temperature contours are superimposed on cutaways of the three-dimensional domains. Volume shaded in red represents the initiation region, and volume shaded in gray represents the reactive region. Dashed pink lines highlight the ignition temperature (here $\theta_\mathrm{ign} = 0.2$). (a) Cylindrical domain with cold wall boundary conditions; (b) cylindrical domain with inert confinement conditions; (c) slab domain with cold wall boundary conditions; (d) slab domain with inert confinement conditions. The black-white dashed lines in (b) and (d) represent the initial interface between the reactive and non-reactive regions.}
\label{Fig_schematic}
\end{figure}

As illustrated in Fig.~\ref{Fig_schematic}(a), the flame was initiated by a high temperature zone (volume shaded red) where the temperature and reaction progress variable were initially $\theta=1$ and $C=0$, respectively; for the reactive medium (unshaded volume), the initial conditions were $\theta=0$ and $C=1$. The thickness and diameter of the reactive medium are denoted as $h$ and $d$, respectively. Only half of the reactive medium needed to be considered by applying an adiabatic boundary condition along the bottom boundary of the domain. The reactive medium loses heat into its surroundings through the top boundary (see Sect.~\ref{Sec2_3}). The numerical method used to solve the continuum formulation was a central difference spatial discretization with the Forward-Euler method used for time integration of $C$.

\subsection{Discrete source model}
\label{Sec2_2}

In the discrete source model, point-like reactive particulates are randomly distributed in a three-dimensional oxidizing environment. The heat released by each particulate diffuses through the inert, inter-particle medium. The mass diffusion of oxidizer is neglected in this model, i.e., $\mathrm{Le} = \infty$, which is consistent with the assumption made in the continuum model. This system is described by a three-dimensional heat diffusion equation with a source term that governs the heat release of the point-like reactive particulates in the system. The non-dimensional form of the governing equation is formulated as follows,
\begin{equation}
\frac{\partial \theta}{\partial t} = \frac{\partial^2 \theta}{\partial x^2} + \frac{\partial^2 \theta}{\partial y^2} + \frac{\partial^2 \theta}{\partial z^2} + \sum\limits_{i=1}^N R_i(\mathrm{\textbf{x}},t)
\label{Eq3}
\end{equation}
where $\mathrm{\textbf{x}}$ is a position vector in three dimensions, i.e., $\mathrm{\textbf{x}} = (x,y,z)$, $N$ is the total number of reactive particles in the system, and $i$ is the particle index. The reaction model for the $i^\mathrm{th}$ particle $R_i(\mathrm{\textbf{x}},t)$ can be expressed as follows,
\begin{equation}
R_i(\mathrm{\textbf{x}},t) = \frac{\delta(\mathrm{\textbf{x}}-\mathrm{\textbf{x}}_i)}{\tau_\mathrm{c}} \mathrm{H} (t - t_{\mathrm{ign},i}) \mathrm{H} (\tau_\mathrm{c} - t + t_{\mathrm{ign},i})
\label{Eq4}
\end{equation}
where $\mathrm{\textbf{x}}_i$ is the location of the $i^\mathrm{th}$ particle, and $t_{\mathrm{ign},i}$ is the time when the local temperature at the $i^\mathrm{th}$ particle first reaches the ignition temperature $\theta_\mathrm{ign}$. The point-like nature of the particle is described by a spatial Dirac $\delta$-function in Eq.~\ref{Eq4}. The Heaviside function $\mathrm{H}$ is used to construct a boxcar function for the cases that the particles release energy over a finite amount of time after being ignited. 

The discreteness parameter $\tau_\mathrm{c}$ is defined as the ratio between the energy release time of each particle $\widetilde{t}_\mathrm{r}$ and the characteristic heat diffusion time over the average inter-particle spacing $\widetilde{l}$, i.e., $\widetilde{t}_\mathrm{d} = \widetilde{l}^2/\widetilde{\alpha}$. The expression for $\tau_\mathrm{c}$ is as follows,
\begin{equation}
\tau_\mathrm{c} = \frac{\widetilde{t}_\mathrm{r}}{\widetilde{t}_\mathrm{d}}
\label{Eq5}
\end{equation}
When $\tau_\mathrm{c} \ll 1$, the spatial discreteness of the heat sources becomes significant. In the limit of $\tau_\mathrm{c} \to 0$, the energy release of each particle is instantaneous, and thus, the source term (Eq.~\ref{Eq4}) becomes $R_i(\mathrm{\textbf{x}},t)= \delta(\mathrm{\textbf{x}}-\mathrm{\textbf{x}}_i) \delta(t - t_{\mathrm{ign},i})$. 

In the discrete source model, time is non-dimensionalized by the characteristic heat diffusion time, i.e., $t = \widetilde{t} / \widetilde{t}_\mathrm{d}$, and spatial coordinates are non-dimensionalized by the average particle spacing, i.e., $x = \widetilde{x} / \widetilde{l}$. Note that this non-dimensionalization is different from that used for the continuum model. Scale conversion between these two dimensionless systems is needed when the results of the discrete source model are compared with those of the continuum model in Sect.~\ref{Sec3}. Further details of how the dimensionless equations are derived from the dimensional governing equations for the discrete source model can be found in Appendix~\ref{SecA1_2} and Ref.~\cite{Lam2017PRE}.

\begin{figure}[h]
\centerline{\includegraphics[width=\textwidth]{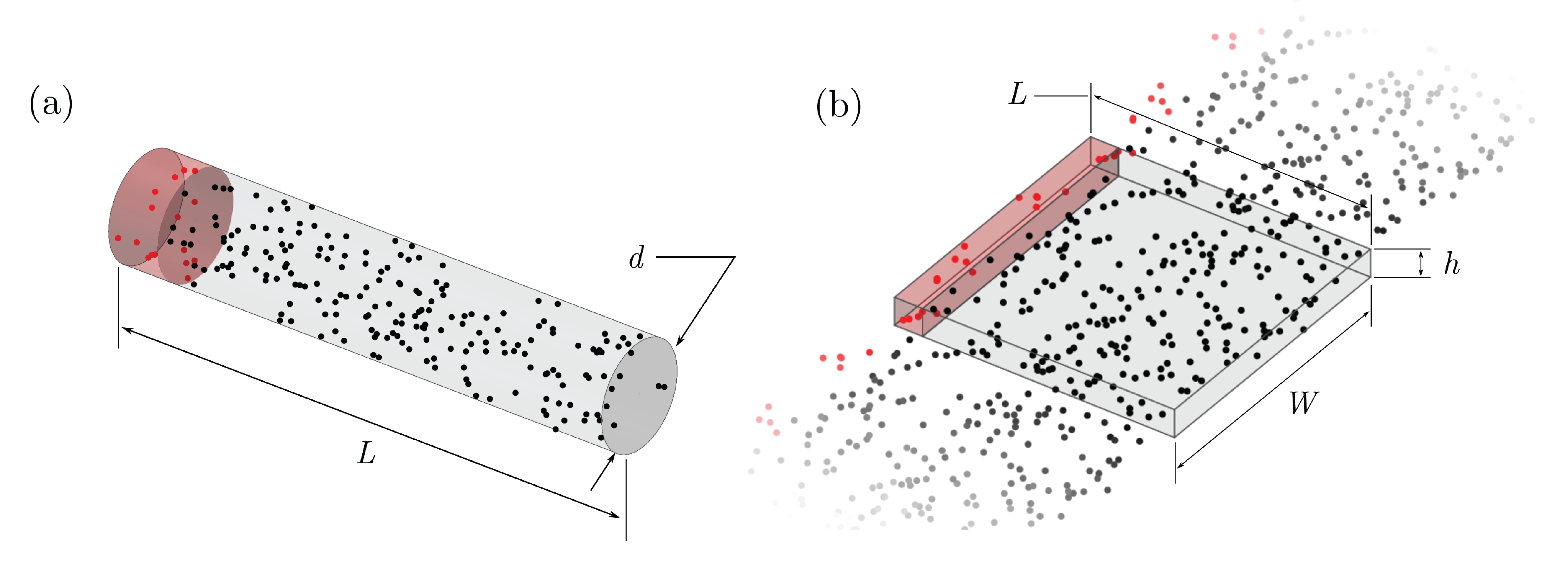}}
\caption{Schematic of discrete model cloud geometries with sources shown as points. Volume shaded in red represents the initiation region, in which point sources are ignited (active) at $t = 0$, but the medium is inert. Point sources in the gray shaded region ignite when the ignition temperature at the specific location is reached. (a) Cylindrical cloud geometry with length $L$ and diameter $d$; (b) slab domain with length $L$, width $W$ and thickness $h$. The implementation of periodic boundary conditions to create an effectively infinite width slab via the method of images is shown schematically using points outside the slab.}
\label{Fig_schematic_discrete}
\end{figure}

As depicted schematically in Fig.~\ref{Fig_schematic_discrete}, a cloud of randomly distributed particulates was initially generated in rectangular slab or cylindrical geometries. The length of the particulate cloud in the flame propagation direction ($x$-direction), $L$, is $10$ times the slab thickness or the cylinder diameter, i.e., $L=10h$ or $L=10d$, respectively. For the slab geometry, the width of the cloud, $W$, equals the cloud length, i.e., $W=L$, so the aspect ratio of the slab is maintained at $h:W=1:10$. Since a periodic boundary condition is applied along the edges of the slab in the $z$-direction (as described in Section 2.3), however, there are no heat losses from the edge, and the slab is thus effectively of infinite width. Within the initiation region (shaded red), the particles (red points) were initially ignited at $t=0$. The length of this initiation region is one tenth of the cloud length, i.e., $L_\mathrm{init}/L=0.1$. The heat generated by these particles triggered the subsequent ignitions and energy release of the particles in the reactive cloud. By solving the governing equations (Eqs.~\ref{Eq3} and \ref{Eq4}), the temperature at a specific location $\theta(\mathrm{\textbf{x}},t)$ can be obtained. This solution for $\theta(\mathrm{\textbf{x}},t)$ can be analytically constructed via linearly superimposing the Green's function solution of an individual particle for $\tau_\mathrm{c}=0$ or its time convolution with a temporal boxcar function described by Eq.~\ref{Eq4}. Shown in Fig.~\ref{Fig_schematic_discrete}(b) are the particle images used to handle the boundary conditions as detailed in Section \ref{Sec2_3}. The details of the computations for the discrete source model can be found in Ref.~\cite{Lam2017PRE}.

The computations are parametrized by the ignition temperature $\theta_\mathrm{ign}$, discreteness parameter $\tau_\mathrm{c}$, and the size of the reactive cloud $h$ or $d$. Concerning the stochastic nature of the randomly distributed particulates, for each set of $\theta_\mathrm{ign}$, $\tau_\mathrm{c}$, and cloud thickness or diameter, simulations were performed for at least $20$ independently generated particulate clouds in order to obtain a statistical ensemble of results. Random particulate positions were generated by the Mersenne Twister pseudorandom number generator (MT19937). \cite{matsumoto1998mersenne} The simulations were executed in parallel on Compute Canada's supercomputers \textit{Mammoth} and \textit{Guillimin}.

\subsection{Losses}
\label{Sec2_3}

The heat losses into the surroundings were implemented in the models via applying boundary conditions to the governing equations. Two types of boundary losses are considered in this study: (1) losses to a cold-wall confinement (Fig.~\ref{Fig_schematic}(a) and (c)); and (2) losses into an essentially infinite amount of cold, inert gas surrounding the reactive medium with heat capacity and conductivity the same as the reactive medium (Fig.~\ref{Fig_schematic}(b) and (d)). The cold-wall boundary condition (1) serves as a model for typically encountered experimental scenarios, where the tube or parallel plates containing metal particulate suspensions are heat sinks that remain at a nearly constant temperature. The inert gas confinement condition (2) would correspond to a reactive suspension dispersed into inert gas or fuel particles dispersed into an oxidizer that does not otherwise contain fuel sources. How these two types of boundary losses are implemented in the continuum and discrete source models are reported in the following paragraphs.

To implement a cold-wall confinement in the continuum model, an isothermal boundary condition was applied along the upper edge of the reactive medium, i.e., $\theta \rvert_{y=h/2}=0$ or $\theta \rvert_{r=d/2} = 0$, as illustrated in Fig.~\ref{Fig_schematic}(a) and (c). The interior of the domain is discretized by a square grid consisting of $N_x$ nodes in the $x$-direction and $N_y$ nodes in the $y$-direction; $\Delta x = \Delta y$ is the grid spacing. The Laplacian $\frac{\partial^2 \theta}{\partial x^2} + \frac{\partial^2 \theta}{\partial y^2}$ in Eqn.~\ref{Eq1} is approximated at each node $(i,j)$ by
\begin{equation}
    \frac{\partial^2 \theta}{\partial x^2} + \frac{\partial^2 \theta}{\partial y^2} \approxeq
    \frac{
        \theta_{i-1,j} + \theta_{i+1,j} + \theta_{i,j-1} + \theta_{i,j+1} - 4\theta_{i,j}
    }{\Delta x^2}
\end{equation}
The boundary condition enters via the Laplacian of the topmost, leftmost and rightmost nodes (i.e., the nodes closest to the isothermal boundary). For the topmost nodes, the Laplacian is approximated by 
\begin{equation}
    \frac{\partial^2 \theta}{\partial x^2} + \frac{\partial^2 \theta}{\partial y^2} \approxeq
    \frac{
        \theta_{i-1,j} + \theta_{i+1,j} - 4\theta_{i,j}
    }{\Delta x^2}
\end{equation}
and similarly for the leftmost and rightmost nodes.
In this case, the computational domain covers only the reactive medium. In order to model an inert-gas confinement, as shown in Fig.~\ref{Fig_schematic}(b) and (d), a reactive medium and a large inert region were included in the computational domain. In an identical manner, an isothermal boundary condition of $\theta = 0$ was applied along the boundary of the computational domain, which was placed sufficiently far away from the edge of the reactive medium so as not to influence the flame.

In the discrete source model (Fig.~\ref{Fig_schematic_discrete}), the condition of inert-gas confinement is imposed by using the Green's function to construct the temperature field. By definition, the free-space Green's function is the solution to the heat equation in an unbounded domain, and any superposition of Green's functions also exhibits the required decay at infinity. Thus, use of the Green's function naturally satisfies the conditions of inert gas confinement outside of the cloud of reactive particles.

The cold-wall confinement condition was imposed by using the well-known method of images.
For the slab geometry, virtual copies (images) of the physical domain are created in both directions perpendicular to the cold-wall boundary with an offset from the physical domain of $2h, 4h, 6h, ...$ . Each image contains virtual particles that contribute to the temperature in the physical domain Then, images of the physical domain mirrored about the cold-wall boundary image are added with an offset from the physical domain of $h, 3h, 5h, ...$ . The contribution of the virtual particles in the latter images are instead subtracted from the temperature in the physical domain.
For the cylindrical geometry, the cold-wall boundary condition is satisfied by a series solution; the details are included in Appendix~\ref{SecA2_1}.

To create a slab with an effectively infinite width in the $z$-direction, periodic boundary conditions are implemented by introducing images of the physical domain as above, but at an offset of $h, 2h, 3h, ...$ with all virtual particles contributing positively. The series solution described in Appendix~\ref{SecA2_2} provides an alternative way to incorporate the periodic boundary condition.


\section{Results}
\label{Sec3}

The results are presented in the following order: for the continuum model, the morphology and flame speed are briefly explored, followed by the problem of critical dimensional scaling. The results for the discrete model are presented in the same order.

\subsection{Continuum model}

In this section, the continuum model non-dimensionalization of length and time quantities (denoted with superscript $*$) is used.

\subsubsection{Continuum morphology and flame speed}

For the continuum model, the temperature field of the reactive medium was directly obtained from numerical simulations. The flame front is defined as the iso-contour of ignition temperature, $\theta_\mathrm{ign}$, that propagates into the unreacted medium. The flame speed was calculated by tracking the flame front position along the central axis of the reactive medium, i.e., the bottom boundary of the computational domain, and then fitting the resulting position-time data, disregarding the first and last 30\% of data in order to reduce initiation effects and the influence of the boundaries normal to the direction of propagation. The flame propagation is said to be successful if the flame front reaches the final 10\% of the length of the domain.

\begin{figure}[h]
\centerline{\includegraphics[width=\textwidth]{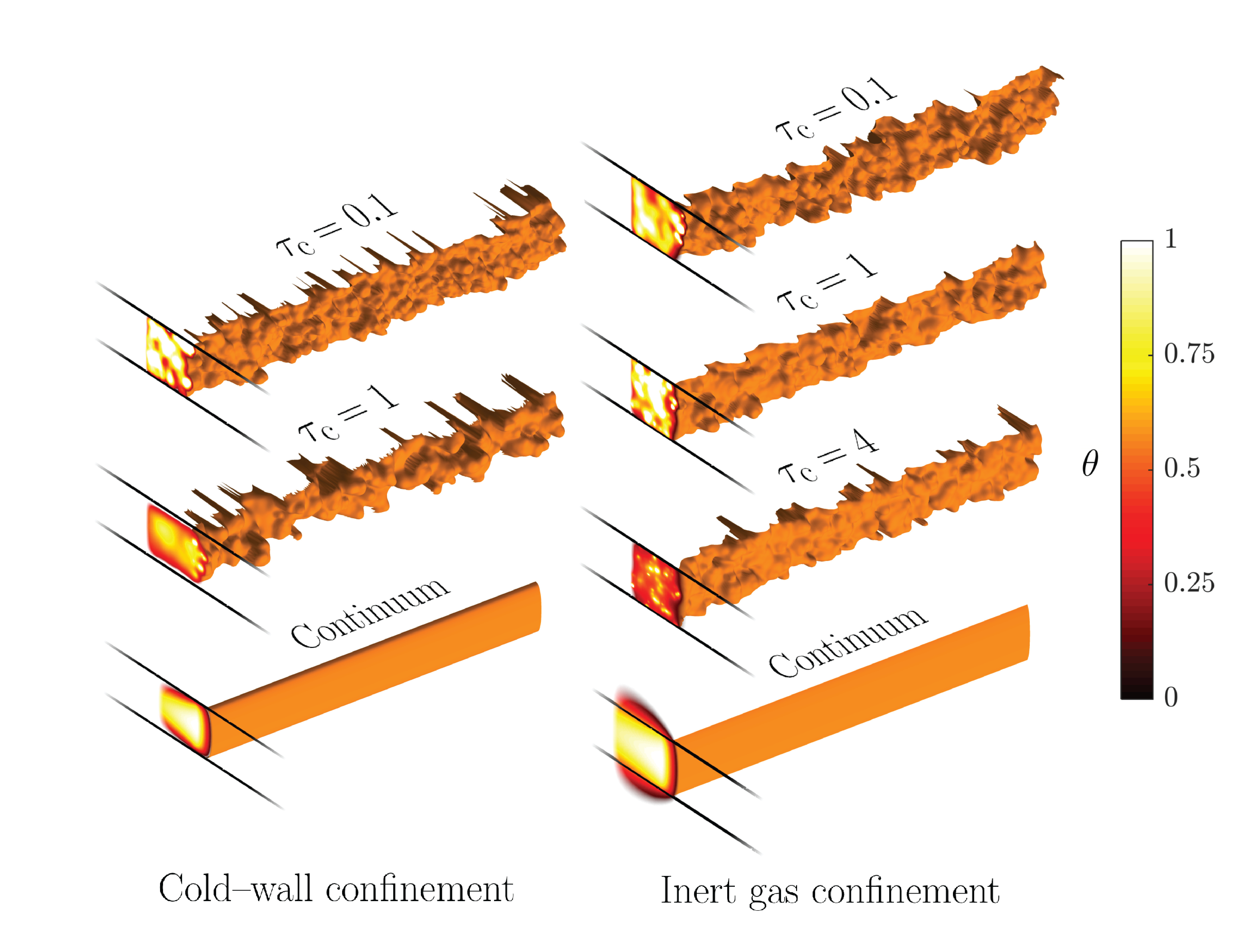}}
\caption{Snapshots of ignition temperature iso-contours of flames propagating to the right for $\theta_\mathrm{ign} = 0.1$ and various values of $\tau_\mathrm{c}$. The foreground corners of the slab geometry are highlighted by solid lines; a two-dimensional slice of the temperature profile is shown as the colored contour plot (corresponding temperature is indicated by the color bar). Each column corresponds to the indicated boundary condition. The continuum flame fronts are produced using the continuum model, while other fronts are produced using the discrete source model.}
\label{Fig_mo}
\end{figure}

The morphology of the flame is exemplified at the bottom of Fig.~\ref{Fig_mo}. Here, a three-dimensional iso-contour of the ignition temperature front ($\theta_\mathrm{ign} = 0.1$) is shown beside colored contours of a two-dimensional slice of the data. Under both boundary conditions (cold-wall, inert gas) the structure is smooth and essentially independent of time.

\begin{figure}[h]
\centerline{\includegraphics[width=\textwidth]{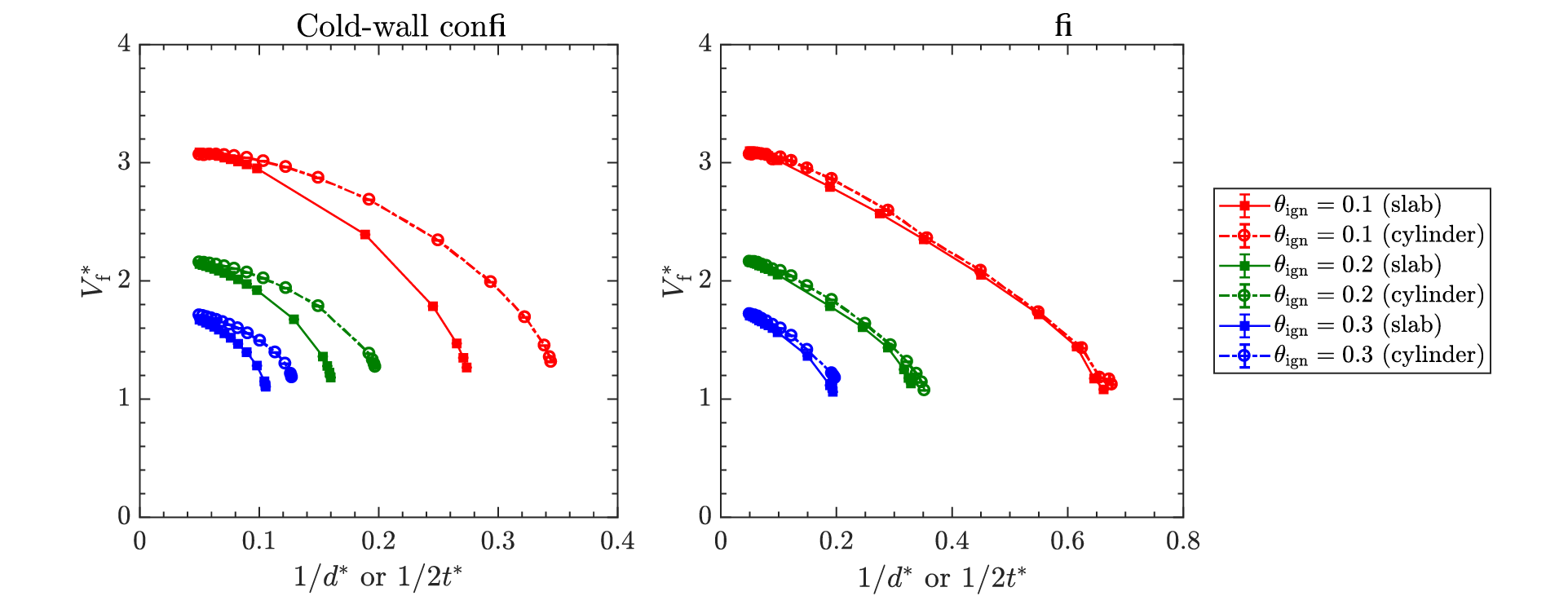}}
\caption{Propagation speed in the continuum non-dimensionalization plotted as a function of the inverse continuum dimension, i.e., (diameter $d^*$ and thickness $h^*$), for both cold-wall confinement and inert gas confinement boundary conditions. Thickness $h$ is scaled by an additional factor of $2$ to illustrate the scaling in the inert gas confinement case.}
\label{Fig_continuum_speed}
\end{figure}

The computed flame speed is plotted as a function of the reciprocal of the continuum dimension of the reactive region in Fig.~\ref{Fig_continuum_speed} for both boundary conditions considered, with ignition temperatures $\theta_\mathrm{ign} = 0.1$, $0.2$, and $0.3$. In the case of inert gas confinement, for a fixed ignition temperature, the propagation speeds are similar between slab and cylinder geometries for $d \sim 2t$.

In Fig.~\ref{Fig_Vf}(a), the solid black curve depicts again the continuum-model results of flame speeds plotted as a function of $1/d^{*}$ for $\theta_\mathrm{ign} = 0.2$. In the limit of $1/d^{*} \to 0$, i.e., for a cylindrical domain with an infinitely large diameter, the continuum-model results approach the one-dimensional analytic solution of flame speed for a continuous reactive medium without losses. The details of this analytic solution can be found in Ref.~\cite{GoroshinLeeShoshin1998}. As diameter decreases ($1/d^{*}$ increases), the resulting $V_\mathrm{f}^{*}$ decreases until reaching a critical diameter ($d_\mathrm{cr}^{*}$ indicated by the vertical gray line in Fig.~\ref{Fig_Vf}(a)) below which flame propagation cannot be sustained.

\subsubsection{Critical dimensions and scaling}

The critical dimension for the continuum model can be defined as the smallest dimension, i.e., the diameter $d^*_\mathrm{cr}$ (for cylinders) or the thickness $h^*_\mathrm{cr}$ (for slabs), of the reactive region that will allow the reactive front to propagate. The critical dimensions for both boundary conditions, both geometries and ignition temperatures $\theta_\mathrm{ign} = 0.1$, $0.2$, and $0.3$ are tabulated in Table~\ref{Table_continuum}. The scaling ratio $d^*_\mathrm{cr} / h^*_\mathrm{cr}$ is computed in the last row.

\begin{table}[h!]
    \centering
    \begin{tabular}{|c||c|c|c||c|c|c|}
        \hline
        & \multicolumn{3}{|c||}{Cold-wall confinement} & \multicolumn{3}{|c|}{Inert gas confinement} \\ \hline
        $\theta_\mathrm{ign}$ & 0.1 & 0.2 & 0.3 & 0.1 & 0.2 & 0.3 \\ \hline
        $d^*_\mathrm{cr}$ & 2.9014 & 5.0617 & 7.8426 & 1.4719 & 2.8777 & 4.9700 \\
        $h^*_\mathrm{cr}$ & 1.8198 & 3.1302 & 4.7440 & 0.7544 & 1.4748 & 2.5411 \\
        $d^*_\mathrm{cr} / h^*_\mathrm{cr}$ & 1.5944 & 1.6170 & 1.6532 & 1.9744 & 1.9512 & 1.9558 \\
        \hline
    \end{tabular}
    \caption{Critical diameter $d^*_\mathrm{cr}$ and thickness $h^*_\mathrm{cr}$, and the scaling ratio $d^*_\mathrm{cr} / h^*_\mathrm{cr}$, for cold-wall and inert gas confinement boundary conditions for various ignition temperatures ($\theta_\mathrm{ign}$) as computed using the continuum model.}
    \label{Table_continuum}
\end{table}

\subsection{Discrete source model}

In this section, quantities obtained from the discrete source model are presented in the dimensionless form for the discrete source system except in Fig.~\ref{Fig_Vf}, where the results of flame speed from the continuum model and the discrete source model are directly compared.

\subsubsection{Morphology of the discrete flame front}

The flame front for the slab geometry is exemplified in Fig.~\ref{Fig_mo} as three-dimensional iso-contours of $\theta = \theta_\mathrm{ign} = 0.1$. In the left column, the cold-wall confinement case for the discrete source model is shown for $\tau_\mathrm{c} = 0.1$ and $\tau_\mathrm{c} = 1$ for a fixed cloud size $h = 8$. For larger $\tau_\mathrm{c}$, the flame fails to propagate at this domain size due to the high rate of heat loss under these boundary conditions. The continuum model flame front is appended at the bottom of the column for comparison.
In the right column of Fig.~\ref{Fig_mo}, the inert gas confinement case is shown for $\tau_\mathrm{c} = 0.1$, $\tau_\mathrm{c} = 1$, and $\tau_\mathrm{c} = 4$, with the continuum model front appended.
For both boundary conditions, as $\tau_\mathrm{c}$ increases, the length scale of the features of the flame front increases and the front becomes smoother. The trailing features in some fronts are due to sharp curvature in the ignition temperature iso-contour (e.g., near the cold-wall boundary). 

Since the discrete source simulations are qualitatively similar for both cold-wall and inert gas confinement cases, the remainder of this study will limit attention to the inert gas confinement case exclusively.

\subsection{Flame speed for particulate clouds}
\label{Sec3_2}

\begin{figure}[h]
\centerline{\includegraphics[width=0.8\textwidth]{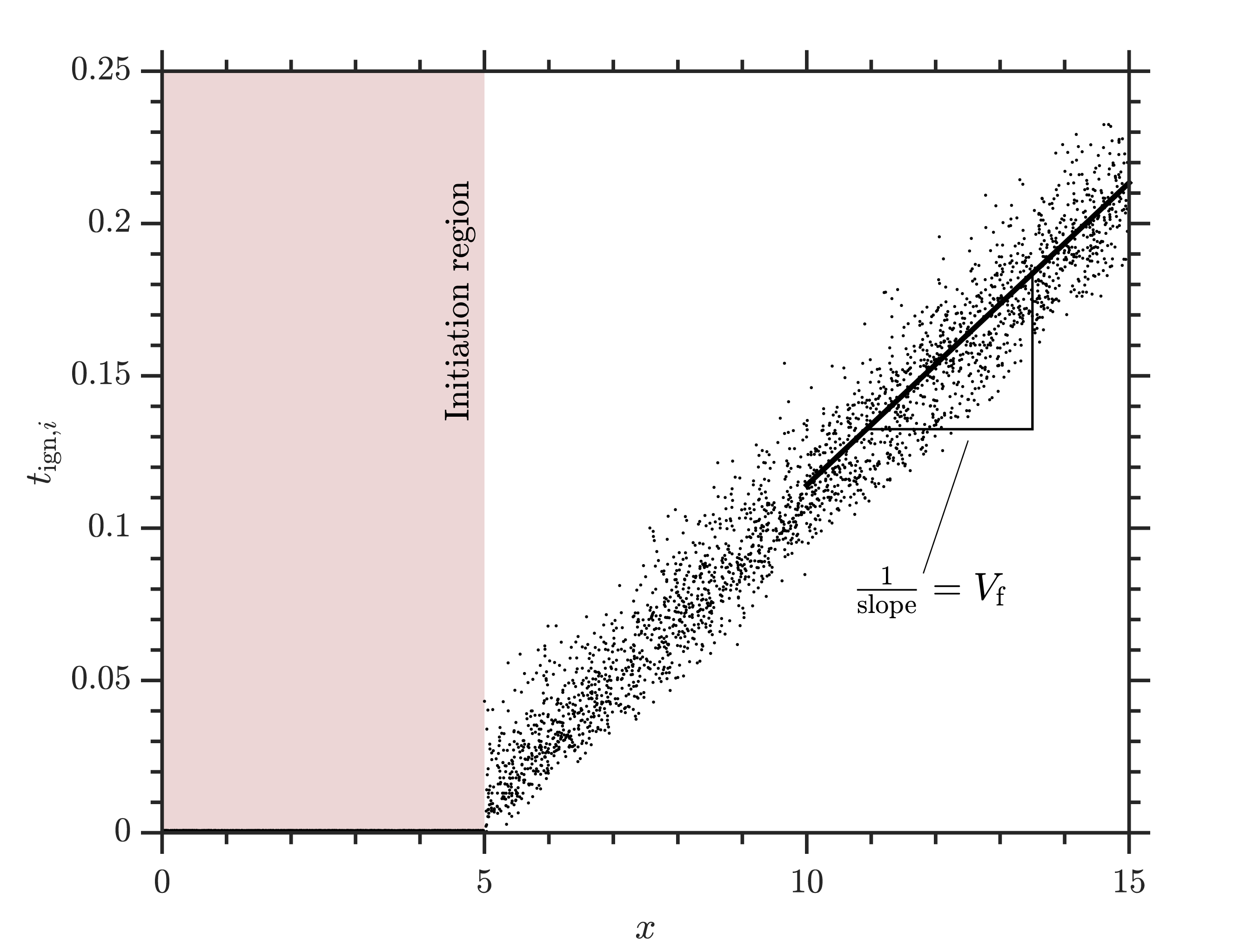}}
\caption{Sequence of source ignition times $t_{\mathrm{ign},i}$ (resulting from the discrete source model) plotted as a function of the corresponding position in the direction of flame propagation ($x$-direction) for a cylindrical cloud with $d=5$, $\tau_\mathrm{c}=0.1$, and $\theta_\mathrm{ign}=0.2$.}
\label{Fig_xt}
\end{figure}

In this model, the results directly obtained from the computations were the sequence of source ignition times, $t_{\mathrm{ign},i}$. Sample results of the sequence of source ignition times for a cylindrical cloud with $d=5$, $\tau_\mathrm{c}=0.1$, and $\theta_\mathrm{ign}=0.2$ are plotted as an $x$-$t$ diagram in Fig.~\ref{Fig_xt}. Within the initiation region on the left, the sources were ignited at $t=0$. Each dot plotted on Fig.~\ref{Fig_xt} represents the ignition time $t_{\mathrm{ign},i}$ of a source and its location in the $x$-direction (i.e., the direction of flame propagation). The temperature field over the entire cloud at specific times was then calculated by evaluating the analytic solution for $\theta(\mathrm{\textbf{x}},t)$ by using the solution of source ignition times over a uniform Cartesian grid (see more details in Ref.~\cite{Lam2017PRE}). The flame speed resulting from the discrete source model was then obtained by fitting the position-ignition-time data of the sources, disregarding $50\%$ of the particles closest to the initiation zone.
In a fashion similar to the continuum analysis, the flame propagation is said to be successful if an ignition is detected in the final $10\%$ of the length of the domain.

\begin{figure}[h]
\centerline{\includegraphics[width=0.9\textwidth]{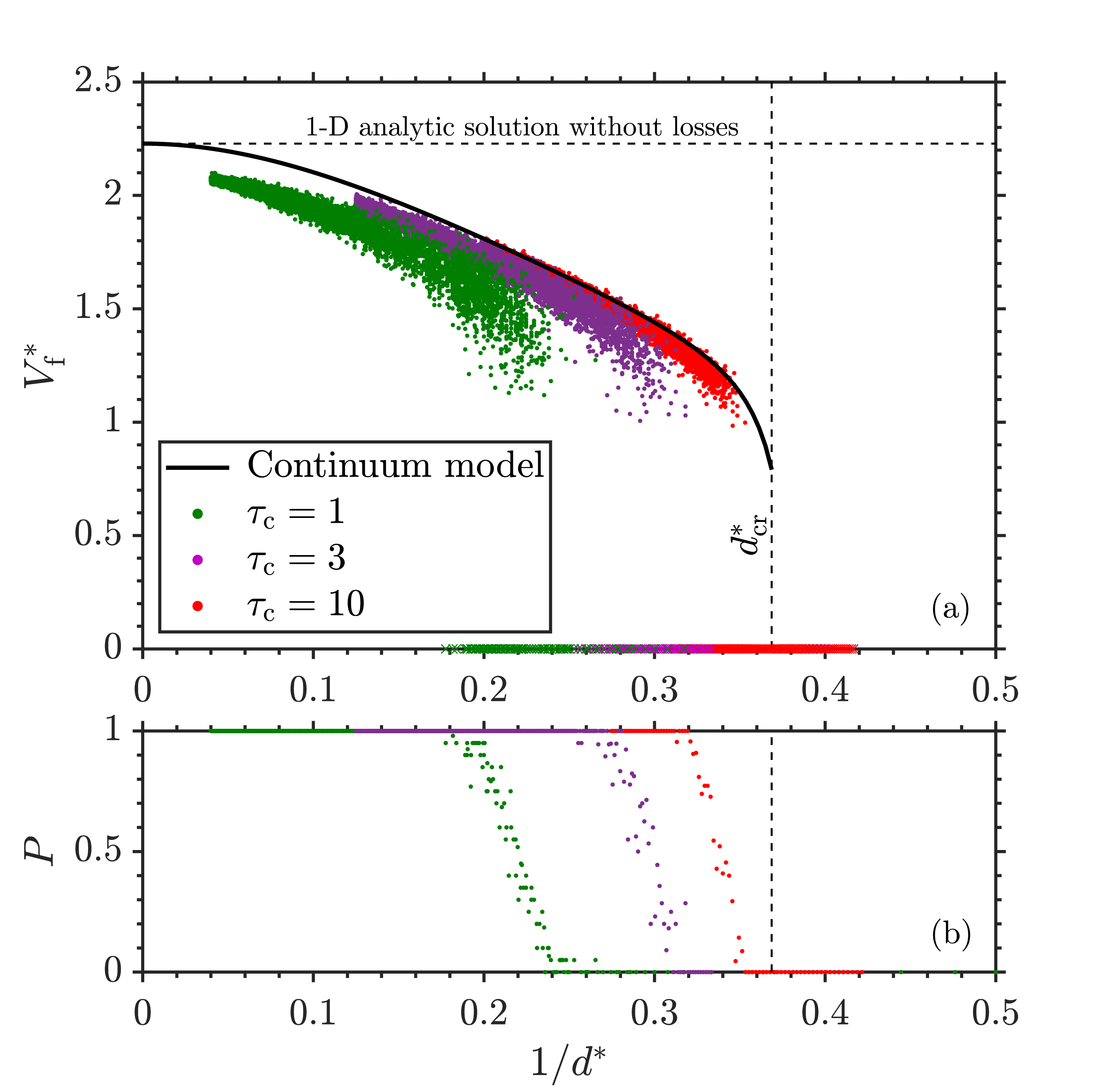}}
\caption{The results of (a) flame speeds $V_\mathrm{f}^{*}$ and (b) propagation probability $P$ as a function of the reciprocal of the reactive cloud diameter $1/d^{*}$ for the cylindrical cases with $\theta_\mathrm{ign}=0.2$ and inert-gas confinement. Horizontal dashed line indicates the flame speed of a one-dimensional analytic solution without losses; vertical dashed line indicates the critical diameter $d^*_\mathrm{cr}$ of the continuum model.}
\label{Fig_Vf}
\end{figure}

The speeds of flame propagation resulting from the continuum model and the discrete source model for selected cylindrical cases with $\theta_\mathrm{ign}=0.2$ and inert-gas confinement are compared in Fig.~\ref{Fig_Vf}(a). Note that the results in this figure are reported as quantities in the continuum-model nondimensionalization, which are denoted with an asterisk $*$. The quantities of flame speed $V_\mathrm{f}^{*}$  and cloud diameter $d^{*}$ in the continuum-model nondimensionalization are related to those in the discrete-source-model nondimensionalization as $V_\mathrm{f}=V_\mathrm{f}^{*}/\sqrt{\tau_\mathrm{c}}$ and $d=d^{*}\sqrt{\tau_\mathrm{c}}$, respectively.

The dots in Fig.~\ref{Fig_Vf}(a) are the time-average flame speeds resulting from the discrete source model plotted as a function of the reciprocal of the cloud diameter. Each dot represents the result from one individual run of the discrete-source-model computations. The results for the cases with $\tau_\mathrm{c}=1$, $3$, and $10$ are plotted as green, purple, and red dots in Fig.~\ref{Fig_Vf}(a), respectively. The markers plotted at $V_\mathrm{f}^{*}=0$ represent the cases where the flame failed to propagate.

As shown in Fig.~\ref{Fig_Vf}(a), for a large value of discreteness parameter, e.g., $\tau_\mathrm{c}=10$, the speed $V_\mathrm{f}^{*}$ resulting from the discrete source model is in agreement with the continuum model results. As $\tau_\mathrm{c}$ decreases, the difference between the results from the discrete source model and the continuum model increases. For $\tau_\mathrm{c}=1$, the speed $V_\mathrm{f}^{*}$ resulting from the discrete source model is significantly less than that resulting from the continuum model at the same cloud diameter.

\subsection{Critical dimensions of particulate clouds}
\label{Sec3_3}

It can be noted in Fig.~\ref{Fig_Vf}(a) that while the continuum model associates each diameter with a fixed propagation speed, the discrete source model exhibits a range of propagation speeds. Furthermore, for each value of $\tau_\mathrm{c}$, at sufficiently small diameters, some realizations of discrete clouds can sustain flame propagation, and some realizations cannot. Thus, for an ensemble of discrete-source-model results at a certain cloud diameter, the probability of flame propagation can be defined as the ratio between the number of runs in which flame propagation was observed, $N_\mathrm{prop}$, and the total number of runs in the ensemble, $N_\mathrm{ens}$, i.e.,
\begin{equation}
P = \frac{N_\mathrm{prop}}{N_\mathrm{ens}}
\end{equation}
following \cite{Tang2009CTM, Tang2011PRE}. The probability of propagation for cylindrical particulate clouds with $\theta_\mathrm{ign}=0.2$ is plotted as a function of $1/d^{*}$ in Fig.~\ref{Fig_Vf}(b). Each dot represents the ensemble of all realizations sharing the same macroscopic parameters. The vertical line showing the continuum model critical diameter is extended from Fig.~\ref{Fig_Vf}(a). To obtain a smooth representation of the probability of propagation, a shifted and scaled error function was fitted to the $P$-vs.-$1/d^{*}$ data. The cloud diameter of $50\%$ propagation probability, $d_{50}$, can be obtained by evaluating the fitting error function at $P=0.5$. Similar analysis was performed for the results of the discrete source model with rectangular slab geometry to obtain the cloud thickness of $50\%$ propagation probability, $h_{50}$. In the rest of this paper, $h_{50}$ and $d_{50}$ are reported as the critical cloud dimensions marking the flame propagation limit. In comparison, the critical dimensions for the continuum model are denoted by $h_\mathrm{cr}$ and $d_\mathrm{cr}$.

\begin{figure}[h]
\centerline{\includegraphics[width=1.0\textwidth]{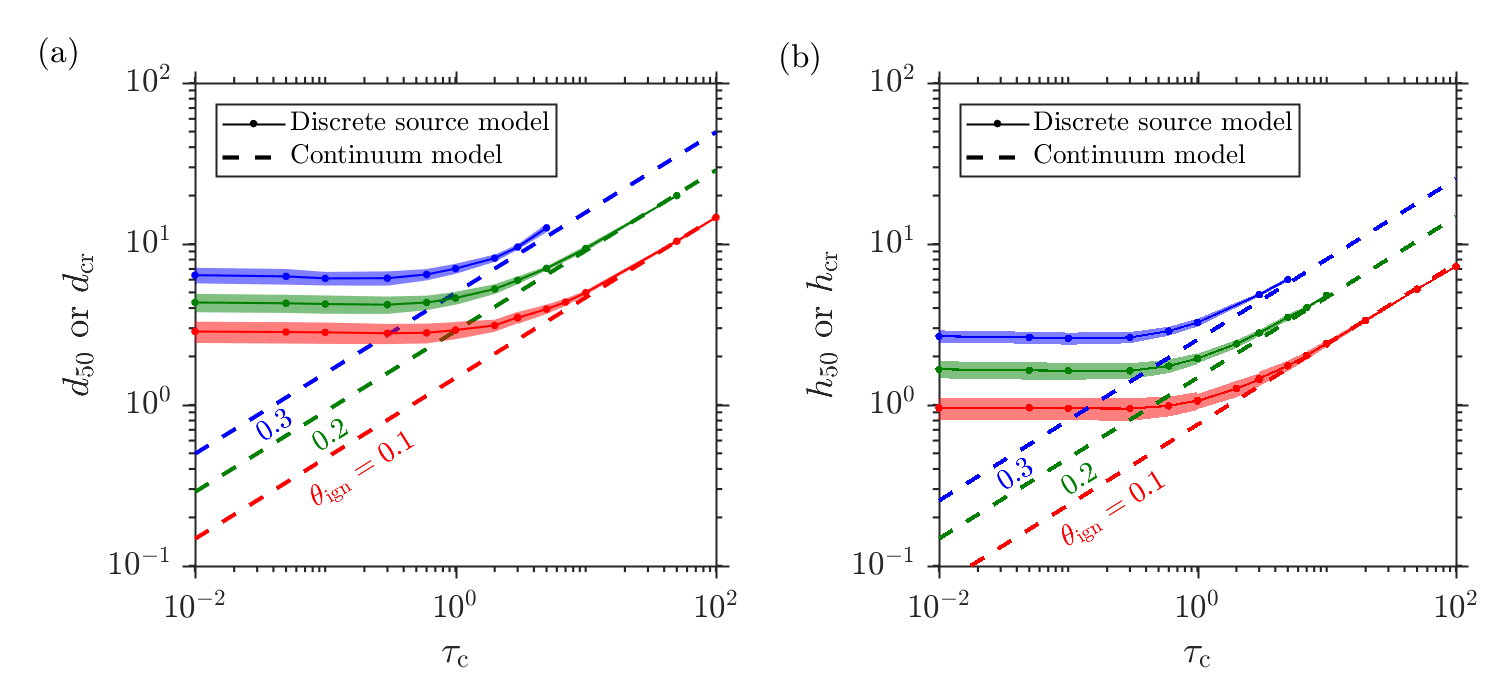}}
\caption{The results of (a) critical cloud thickness ($h_\mathrm{cr}$ or $h_{50}$) for rectangular slab geometry and (b) critical cloud diameter ($d_\mathrm{cr}$ or $d_{50}$) for cylindrical geometry plotted as a function of discreteness parameter $\tau_\mathrm{c}$ on a logarithmic scale. The dots represent the discrete model results of cloud thickness or diameter of $50\%$ propagation probability ($P=0.5$). The upper and lower bounds of the shaded regions enclosing the dots represent the cloud thickness or diameter of $P=0.9$ and $P=0.1$, respectively.} 
\label{Fig_cr}
\end{figure}

In Fig.~\ref{Fig_cr}(a) and (b), the critical dimensions resulting from the continuum model (dashed lines) and the discrete source model (dots connected by solid curves) are plotted as a function of the discreteness parameter $\tau_\mathrm{c}$ for rectangular slab and cylindrical geometries, respectively. The upper and lower bounds of the shaded regions enclosing the results of the discrete source model represent the cloud thickness or diameter of $P=0.9$ and $P=0.1$, respectively. The results for the cases with three different values of $\theta_\mathrm{ign}$, i.e., $\theta_\mathrm{ign}=0.1$, $0.2$, and $0.3$, are compared in this figure. For the same value of $\tau_\mathrm{c}$, the resulting critical dimension increases for increasingly large $\theta_\mathrm{ign}$.

As shown in Fig.~\ref{Fig_cr}, in the limit of large $\tau_\mathrm{c}$ for both geometries, the critical dimensions resulting from the continuum and discrete source models are in agreement. As $\tau_\mathrm{c}$ decreases, the discrete-source-model results deviate from the continuum-model results for the same $\theta_\mathrm{ign}$. In the limit of $\tau_\mathrm{c} \to 0$, the critical dimensions resulting from the discrete source model for each $\theta_\mathrm{ign}$ approach a plateau value.

\subsection{Dimensional scaling between cylindrical and slab geometries}
\label{Sec3_4}

\begin{figure}[h]
\centerline{\includegraphics[width=0.8\textwidth]{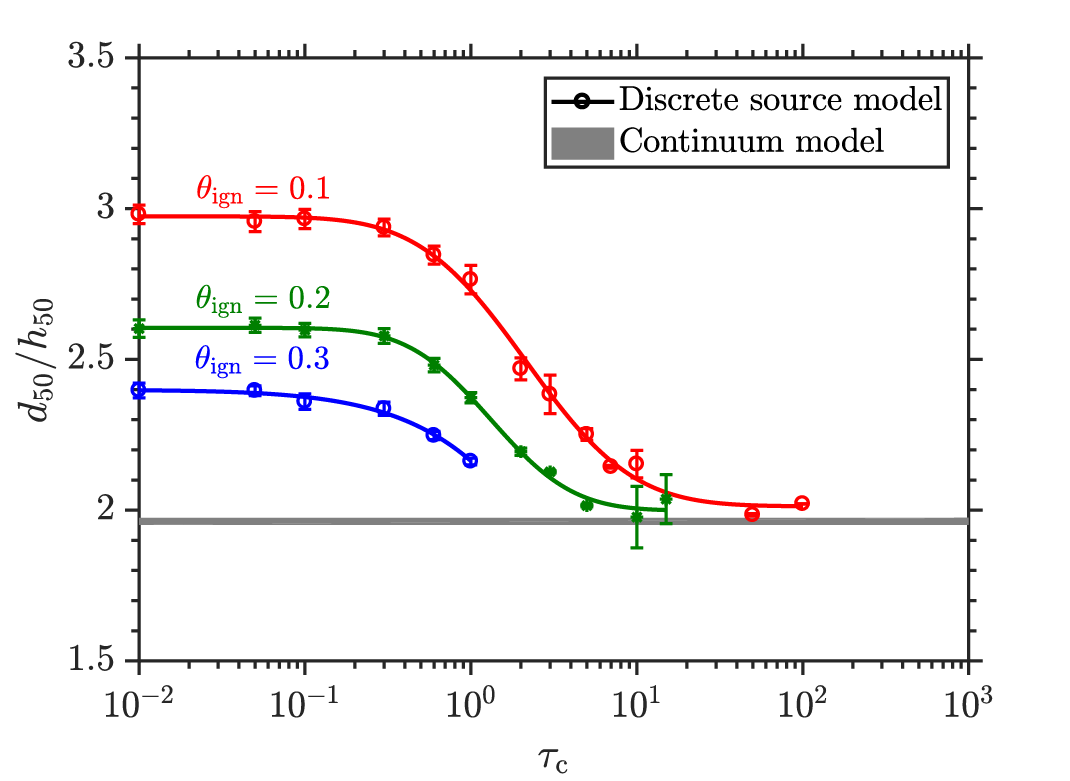}}
\caption{Scaling ratio between critical diameter and critical thickness at $50\%$ probability of propagation plotted as a function of discreteness parameter $\tau_\mathrm{c}$ with horizontal axis plotted on a logarithmic scale. The horizontal gray band indicates the range of values of scaling ratio obtained from the continuum model. Error bars represent the combined root-mean-square error of the propagation probability fit.} 
\label{Fig_sc}
\end{figure}

Comparison of the critical diameters and thicknesses resulting from the discrete source model are obtained through evaluation of the scaling ratio at $50\%$ propagation probability, i.e., $d_{50}/h_{50}$. The scaling ratio is plotted as a function of $\tau_\mathrm{c}$ using logarithmic scales in Fig.~\ref{Fig_sc}. The size of error bars above and below data points are given by the expression
\begin{equation}
\epsilon =
\sqrt{\epsilon_\mathrm{rms,cyl}^2 + \epsilon_\mathrm{rms,slab}^2}
\end{equation}
where each $\epsilon_\mathrm{rms}$ is the root-mean-square error of the fit of probability of propagation as a function of $1/d$ or $1/h$. The depicted curves are scaled and shifted error function fits.
In the limit of large $\tau_\mathrm{c}$, the scaling ratios resulting from the discrete source model with various $\theta_\mathrm{ign}$ tend to approach the scaling ratio near $d_\mathrm{cr}/h_\mathrm{cr}=2:1$. The range of scaling ratios resulting from the continuum model for $\theta_\mathrm{ign} = 0.1$, $0.2$, and $0.3$ listed in Table~\ref{Table_continuum} are depicted as a horizontal shaded band in Fig.~\ref{Fig_sc}.

The scaling ratio in the discrete source model $d_{50}/h_{50}$ increases as $\tau_\mathrm{c}$ decreases. For $\tau_\mathrm{c} \leq 0.5$, $d_{50}/h_{50}$ approaches a plateau value for each $\theta_\mathrm{ign}$. The plateau value of $d_{50}/h_{50}$ in the limit of $\tau_\mathrm{c} \to 0$ decreases for increasingly large $\theta_\mathrm{ign}$.


\section{Discussion}
\label{Sec4}

\subsection{Interpretation of the results in the limit of $\tau_\mathrm{c} \to \infty$}

In the limit of $\tau_\mathrm{c} \to \infty$, the time scale over which an ignited particulate deposits its energy into the medium is much greater than the characteristic time of heat diffusion from one particulate to its closest neighbors. Under such a regime, the spatial discreteness imposed by the random distribution of point sources is homogenized by heat diffusion during this long heat release process. As shown in Fig.~\ref{Fig_mo}, for an increasingly large $\tau_\mathrm{c}$, the resulting flame front exhibits less roughness. In the limit of $\tau_\mathrm{c} \to \infty$, the resulting flame exhibits a laminar-like structure, featuring an overall smooth flame front, which resembles that resulting from the continuum model. The similar speeds of propagation between particulate clouds with large $\tau_\mathrm{c}$ and the continuum model cylinders in Fig.~\ref{Fig_Vf}(a) further confirm the continuum-like propagation of such clouds despite the discreteness of the heat sources.


As shown in Figs.~\ref{Fig_cr} and \ref{Fig_sc}, for large values of $\tau_\mathrm{c}$, the critical cloud thickness and diameter at $50\%$ propagation probability and the scaling ratio between cylindrical and rectangular slab geometries $d_{50}/h_{50}$ resulting from the discrete source model are in agreement with the continuum-model results obtained via finite-difference numerical simulations. This agreement suggests that there is fundamentally no difference in the propagation and quenching mechanisms between a thermally homogenized flame in a particulate cloud and a laminar flame in a continuous medium subject to boundary heat losses.


\subsection{Interpretation of the results in the limit of $\tau_\mathrm{c} \to 0$}

The limit of $\tau_\mathrm{c} \to 0$ represents a regime where propagation is limited by heat diffusion between sources. The propagation speed as depicted in Fig.~\ref{Fig_Vf}(a), is consequently lower than that of the continuum model. The particular distribution of sources further affects the propagation velocity, especially near the critical dimension of the cloud. Propagation through sufficiently small clouds may typically rely on ``paths" traced by sequences of neighboring particles that bottleneck the speed of propagation. The absence of these paths in small cylinders may lead to quenching, while the abundance of these paths in large cylinders allow more consistent propagation speeds. 

The propagation probability shown in Fig.~\ref{Fig_Vf}(b) varies over a larger range of $1/d^*$ for the case $\tau_\mathrm{c} = 1$, with $P$ strongly dependent on $d^*$ for values of $d^*$ between the range of $4.2$ and $5.5$. As $\tau_\mathrm{c}$ increases, the range over which $P$ changes decreases; for $\tau_\mathrm{c} = 50$, propagation probability changes for $d^*$ between $2.8$ and $3.1$, approaching the sharply defined critical dimension found in the continuum model. An alternative illustration is provided by the shaded regions in Fig.~\ref{Fig_cr} that collapse onto a single curve as $\tau_\mathrm{c}$ increases.

The stochastic nature of propagation through clouds of small $\tau_\mathrm{c}$ enhances the importance of cloud geometry. The scaling ratio depicted in Fig.~\ref{Fig_sc} is a function of $\tau_\mathrm{c}$, and varies most significantly from the continuum model value when the ignition temperature is small, i.e., when the mechanism of propagation is the delivery of heat from a small number of nearby sources without allowing the medium to thermally equilibrate.

The question arises as to whether the flame propagation in the limit of this highly discrete regime (i.e., $\tau_\mathrm{c} \to 0$) can be described by percolation theory---a sub-discipline of statistical mechanics that describes how random connections between elements can form a domain-spanning cluster. Researchers have attempted to use percolation theory to interpret combustion phenomena in heterogeneous reactive media with randomly distributed sources (or the so-called disordered media in the literature of statistical mechanics) since the 1980s. \cite{Albinet1986} Beer and Enting, perhaps for the first time, pointed out that a simple site-bond percolation model---considering that a source can only be ignited by its nearest neighbors---cannot adequately describe flame propagation phenomena in such a system wherein the ignition of an individual source might be a result of the collective heat contribution from ignited sources located at distances that are much greater than the average source spacing. \cite{Beer1990} As recently shown by Schiulaz \textit{et al}., site-bond percolation theory is applicable to a system wherein strong volumetric heat loss is present so that the heat contribution to ignite a source is limited within its closest neighbors. \cite{Schiulaz2017,schiulaz2018theory}
It is rather clear that a simple site-bond percolation theory is inadequate to describe the current results of flame propagation in a highly discrete regime. As shown by Grinchuk and Rabinovich, a modified percolation model that considers heating from previously ignited, non-adjacent sources better agrees with their simulation results of flame propagation in a heterogeneous medium. \cite{Grinchuk2004} Thus, future efforts should be made to explore the use of long-range percolation models wherein very distant elements can be connected in order to describe the herein identified discrete combustion phenomena. \cite{schulman1983long}

Recently, using a similar model and semi-analytic solution, Bozorg \textit{et al}. investigated the critical behavior of flame propagation in a suspension of iron particles between two parallel cooling plates. \cite{Bozorg2019} These authors have demonstrated that, as the spacing between the two plates or the temperature of the plates increases, the propagation speed of the flame increases. These findings are consistent with the well-known behavior of laminar flames, thus are inconclusive in revealing the nature of discrete flame. In the current study, via examining the scaling ratio of the critical dimensions for two different geometries, the difference in the fundamental mechanisms underlying continuum and discrete regimes of flame propagation has been demonstrated. Therefore, in order to gain further insights into discrete combustion, it is of importance to select and gauge the metrics that reflect the unique physics underlying such a combustion regime.

\subsection{Implications for experiments}
Because of the central role that flame quenching plays in both combustion theory and applications---particularly in combustion safety---flame quenching in various geometries has been extensively studied. For laminar flames in premixed gas, results of quenching dimensions can be quantitatively scaled between various geometries (quenching plates, tubes and tube bundles, honeycomb structures, packed beds, random porous media, etc.). \cite{Berlad1955, Hackert1998, diMare2000, Joo2006}  The results of the present study are consistent with this picture:  For the continuum model considered in this study, there exists a straightforward, approximate 2:1 scaling between results of flames in cylindrical and slab geometries for the light confinement imposed by the surrounding inert gas, and a scaling ratio somewhat below 2:1 for cold-wall confinement.  When the flame propagation is influenced by the random nature of the discrete, particulate sources, however, the simple 2:1 geometric scaling is broken, as illustrated in Fig.~\ref{Fig_sc}.  In a high aspect ratio channel with a low ignition temperature fuel, for example, a flame is able to continue propagation in a much narrower channel than would be predicted from 2:1 scaling from quenching results of the same mixture in a cylindrical tube.  This result is attributed to the flame's ability to exploit fluctuations in concentration (i.e., local inter-particle spacing) that exist across the channel to continue its propagation under conditions where it would quench in a cylinder.
To date, the studies that have examined flame quenching in suspensions of nonvolatile fuels have used channels formed by quenching plates that would correspond to the case of slab geometry with cold walls considered in the present study. \cite{JAROSINSKI19881917, GOROSHIN1996147, TANG20091905, TANG20111975, PALECKA20152463} The results of our study suggest that the geometry of the quenching tests may have a significant influence on the results---an influence that cannot be accounted for by a fixed scaling parameter in the discrete regime. This conclusion has implications for explosion hazard assessment of particulate fuels.


\section{Concluding remarks}
\label{Sec5}
The present study investigated the decrease in flame speed, leading to flame quenching, in a medium governed by a switch-type reaction mechanism representative of particulate fuels in suspension in a gaseous oxidizer.  Flame propagation in two different geometries was considered, flames in cylindrical columns and rectangular slabs, and two different mechanisms of heat loss were implemented, isothermal walls and confinement by inert gas. In a continuum-based model, which is valid when the particulate spacing is negligible compared to the thermal thickness, the reactive diffusion equation was solved via conventional finite-difference calculations.  In this regime, the scaling between the cylindrical and rectangular (slab) geometries has a unique value of $d_\mathrm{cr}/h_\mathrm{cr} \approxeq 2$, independent of the other parameters in the model.  In a discrete-source model that explicitly treats the spatially discrete nature of the heat sources, the solution was found via superposition of the analytic solution for individual sources (Green's function).  In the limit of fast combustion time in comparison to interparticle diffusion time, the resulting flame propagation became inherently statistical, necessitating ensemble averaging of results generated by parallel supercomputer simulations.  In this discrete regime, a unique value of scaling between the cylindrical and rectangular (slab) geometries was not found, with the scaling values being $d_\mathrm{cr}/h_\mathrm{cr} > 2$ and dependent upon the ignition temperature of the sources.  As the characteristic combustion time was increased, the results of the discrete simulations reverted back to the classical geometric scaling of $d_\mathrm{cr}/h_\mathrm{cr} \rightarrow 2$. The finding of a break-down in simple geometric scaling between the different geometries (cylinder vs. slab) is proposed as a signature of the influence of spatially discrete effects on flame propagation and should be observed experimentally.


\section*{Acknowledgements}

This project was supported under the Canadian Space Agency Flights and Fieldwork for the Advancement of Science and Technology (FAST) funded project, ``Sounding Rocket Flight to Explore Percolating Reactive Waves.'' Computing resources used in this work were provided by Compute Canada. 
The authors are grateful to F.D. Tang and C. Wagner for their contributions to the early stages of this study, and thank S. Goroshin, J.M. Bergthorson, and J. Palecka for valuable discussions and feedback.


\appendix

\section{Derivation of the dimensionless form of the governing equations from the reactive Navier-Stokes equations}
\label{SecA1}

\subsection{Continuum model}
\label{SecA1_1}

In the absence of viscosity and body forces, the full, dimensional model for a single-step reaction based on the conservation laws can be written similarly to \cite{Law2010book} as
\begin{equation}
\label{eqn:dimensionalconservation}
\left \{
\begin{array}{lll}
    \frac{\partial \tilde{\rho}}{\partial \tilde{t}} + \nabla \cdot (\tilde{\rho} \tilde{\mathbf{u}}) = 0 \\
    \frac{\partial \tilde{\mathbf{u}}}{\partial \tilde{t}} + (\tilde{\mathbf{u}} \cdot \nabla) \tilde{\mathbf{u}} = -\frac{\nabla \tilde{p}} {\tilde{\rho}} \\
    \tilde{\rho} \frac{\partial \tilde{h}}{\partial \tilde{t}} + \tilde{\rho} (\mathbf{\tilde{u}} \cdot \nabla) \tilde{h} = \frac{\partial \tilde{p}}{\partial \tilde{t}} + 
    \nabla \cdot \left( \frac{\tilde{k}}{\tilde{c}_p} \nabla \tilde{h}  \right) + \tilde{q}_{\mathrm{c}}\tilde{B} \tilde{\omega} \\
    \tilde{p} = \tilde{p} (\tilde{\rho}, \tilde{T})
\end{array}
\right.
\end{equation}
where dimensional quantities present are the density $\tilde{\rho}$, time $\tilde{t}$, fluid velocity $\tilde{\mathbf{u}}$, fluid pressure $\tilde{p}$, enthalpy $\tilde{h}$, fluid conductivity $\tilde{k}$, heat capacity $\tilde{c}_p$, heat of combustion per mass $\tilde{q}_\mathrm{c}$, fuel per unit volume $\tilde{B}$, and reaction rate $\tilde{\omega}$.

When the pressure fluctuation term $\frac{\partial \tilde{p}}{\partial \tilde{t}}$ is small, the energy equation can be approximated by
\begin{equation}
\tilde{\rho} \tilde{c}_p \frac{\partial \tilde{T}}{\partial \tilde{t}} + \tilde{\rho} (\mathbf{\tilde{u}} \cdot \nabla) \tilde{h} = 
    \nabla \cdot \left( \tilde{k} \nabla \tilde{T}  \right) + \tilde{q}_{\mathrm{c}}\tilde{B} \tilde{\omega}
\end{equation}
where $\tilde{T}$ is the temperature of the fluid. For constant thermophysical properties, one may choose the non-dimensionalization described in Section \ref{Sec2_1} and replace the dimensional terms with non-dimensional terms

\begin{equation}
    \frac{\partial \theta}{\partial t} = - \mathrm{Pe}~\mathbf{u} \cdot \nabla \theta + \nabla^2 \theta + R
\end{equation}
where $\mathrm{Pe} = \tilde{U}_\mathrm{char} \tilde{l} / \tilde{\alpha}$ is the Peclet number based on a characteristic flow velocity $\tilde{U}_\mathrm{char}$, characteristic length $\tilde{l}$ and heat diffusivity $\tilde{\alpha}$. Neglecting the effects of flow induced by thermal expansion of the fluid, one may discard the advection term and examine only the reaction-diffusion equation
\begin{equation}
\label{eqn:rdequationT}
    \frac{\partial \theta}{\partial t} = \nabla^2 \theta + R
\end{equation}

Similarly, the concentration of the limiting reagent (fuel if lean, oxidizer if rich) can be non-dimensionalized as $C = \tilde{C}/\tilde{C}_0$ where $\tilde{C}$ is the dimensional quantity, and $\tilde{C}_0$ is the initial oxidizer concentration. The governing equation then becomes
\begin{equation}
\label{eqn:rdequationC}
    \frac{\partial C}{\partial t} = \nabla^2 C - R.
\end{equation}
A switch-type kinetic model can then be introduced to describe $R$ as in Section \ref{Sec2_1}.

\subsection{Discrete source model}
\label{SecA1_2}

The discrete source model relies on the same set of dimensional equations as the continuum model, but with length non-dimensionalized by the inter-particle spacing and time by the characteristic heat diffusion time $\tilde{t}_\mathrm{d}$ so that, on average, a unit volume contains one discrete source. The reaction term in Eq.~\ref{Eq3} can model a cloud of discrete sources by multiplying the reaction rate by the Dirac $\delta$, i.e., $\delta(\mathbf{x}-\mathbf{x}_i)$ for the $i^\mathrm{th}$ source, and summing over all particles. The total amount of energy released is equivalent to the continuum model in the sense that every unit volume contains, on average, unit energy.


\section{Implementation of boundary conditions in the discrete source model}
\label{SecA2}

\subsection{Series solution for a cold-wall cylinder}
\label{SecA2_1}

A solution for Eq.~\ref{Eq3} for a single particle placed somewhere within a cylinder of radius $b$ representing a cold wall (Dirichlet) boundary condition can be assumed to have the form
\begin{equation}
    \theta(x,r,\phi,t) = \psi_x(x,t) \psi_{r,\phi}(r,\phi,t).
\end{equation}
Here $\psi_x$ is the one-dimensional Green's function for the $x$-direction (boundary conditions at infinity). The coupling of the polar coordinates $\phi$ and $r$ require these variables to be present in the same factor $\psi_{r,\phi}$.

The factor $\psi_{r,\phi}$ can be expressed by a summation of eigenfunctions of the disk; that is,
\begin{equation}
    \psi_{r,\phi} =
    \frac{1}{\pi b^2}
    \sum_{m=0}^{\infty}
    (2-\delta_{0m})
    \cos\left( m (\phi-\phi_0) \right)
    \sum_{n=1}^{\infty}
    \frac{
        J_m\left(
        \frac{\alpha_{mn} r_0}{b}
        \right)
        J_m\left(
        \frac{\alpha_{mn} r}{b}
        \right)
    }{
        \left[ J_{m+1} (\alpha_{mn}) \right]^2
    }
    \mathrm{e} ^{
        -\frac{\alpha_{mn}^2 t}{b^2}}
\label{eqn:isothermalcylsoln}
\end{equation}
where $\delta_{ij}$ is the Kronecker delta; $(r_0,\phi_0)$ are the coordinates of the particle; $J_\nu$ is the Bessel function of the first kind; and $\alpha_{mn}$ is the $n^\mathrm{th}$ zero of $J_m$. This analytic solution is documented in the literature---for example, in \cite{thambynayagam2011diffusion}.

\subsection{Series solution for a periodic slab domain}
\label{SecA2_2}

In this subsection we construct the series solution of Eq.~\ref{Eq3} for a single particle placed at the origin of a three-dimensional domain for which the direction of propagation has boundary conditions applied at infinity (i.e., inert gas confinement) and the transverse directions have periodic boundary conditions. $\tau_\mathrm{c}$ is assumed to be zero; the solution for any value of $\tau_\mathrm{c} > 0$ can be derived by a subsequent convolution operation.

Assuming that the solution has the form
\begin{equation}
    \theta(x,y,z,t) = \psi_x(x,t) \psi_y(y,t) \psi_z(z,t)
\end{equation}
one deduces that the factors $\psi$ required are the one-dimensional Green's functions for the appropriate boundary conditions applied; that is, the free-space Green's function
\begin{equation}
    \psi_x(x,t) = \frac{1}{(4\pi t)^\frac{1}{2}} \exp \left( \frac{-x^2}{4t} \right)
\end{equation}
and the Green's function for a periodic domain of width $L$
\begin{equation}
    \psi_y(y,t) =
    \frac{1}{L}
    \left (
        1+2\sum_{k=1}^{\infty}
            \cos \left (
                \frac{2\pi ky}{L}
            \right )
            \exp \left(
                -\frac{4\pi^2k^2 t}{L^2}
            \right )
    \right )
\end{equation}
and similarly for $\phi_z$. While both the series solution and the method of images for bounded domains require a theoretically infinite number of computations, the method of images is preferable when considering short times $t$ after source ignition, since for small $t$ the Green's function decays rapidly in space, allowing computations to be truncated after a small number of images; the same cannot be said of the series solution, which may be preferable when solutions at large $t$ are needed. \cite{haberman1983elementary}

\subsection{Series solution for a cold-wall slab}
\label{SecA2_3}

The series solution for a cold-wall slab can be derived in a similar fashion to constructing the series solution for a periodic domain described in Appendix~\ref{SecA2_2}. In this case, the one-dimensional Green's function $\psi_z$ for a single source placed at position $z$ in a domain of length $L$ is given by
\begin{equation}
    \psi_z(z,t) = \frac{2}{L} \sum_{k=1}^{\infty} \sin \left( \frac{k \pi y}{L} \right) \sin \left( \frac{k \pi y_0}{L} \right) \exp \left( -\frac{k^2 \pi^2 t}{L^2} \right)
\end{equation}
is used to construct the solution $\theta = \psi_x (x,t) \psi_y (y,t) \psi_z (z,t)$, where each $\psi$ is chosen according to the boundary condition.

%

\bibliographystyle{tfq}
\bibliography{Flame}

\end{document}